\documentclass[paper]{JHEP3}
\usepackage{cite}
\usepackage{amssymb,amsmath}
\usepackage{axodraw4j}		
\usepackage{graphicx}
\usepackage{subfigure}
\graphicspath{{figures/}}

\newcommand{\beq}{\begin{equation}}
\newcommand{\eeq}{\end{equation}}
\newcommand{\beqa}{\begin{eqnarray}}
\newcommand{\eeqa}{\end{eqnarray}}
\newcommand{\cm}{{\cal M}^0}

\newcommand{\wt}{\widetilde}

\newcommand{\order}[1]{{\cal O}(#1)}
\newcommand{\q}[1]{#1_q}
\newcommand{\qb}[1]{#1_{\bar{q}}}

\newcommand{\Q}[1]{#1_Q}

\newcommand{\qpi}[1]{\hat{#1}_{q'}}

\newcommand{\gli}[1]{\hat{#1}_g}

\newcommand{\gaussf}[4]{\ensuremath{\, _2F_1 \left(#1,#2,#3;#4\right)}}

\def\e{\epsilon}

\def\d{{\rm d}}
\def\JET{J}

\newcommand{\mycomment}[1]{ }

\preprint{
TTK-12-41\\				
LPN12-105 \\                       
}

\title{
Antenna subtraction with massive fermions at NNLO: Double real initial-final
configurations
} 
\author{Gabriel Abelof, Aude Gehrmann-De Ridder\\ Institute for Theoretical Physics, ETH, CH-8093 Z\"urich, Switzerland}
\author{Oliver Dekkers\\ Institut f\"ur Theoretische Teilchenphysik und Kosmologie, RWTH Aachen University, D-52056 Aachen, Germany}


\abstract{We derive the integrated forms of specific initial-final tree-level four-parton antenna functions involving a massless initial-state parton and a massive final-state fermion as hard radiators. These antennae are needed in the subtraction terms required to evaluate the double real corrections to $t\bar{t}$ hadronic production at the NNLO level stemming from the partonic processes $q\bar{q}\to t\bar{t}q'\bar{q}'$ and $gg\to t\bar{t}q\bar{q}$.}

\begin{document}
\bibliographystyle{JHEP-2}
\allowdisplaybreaks

\section{Introduction}
Top-quark pair production is one of the key measurements performed at the LHC. The detailed analysis of the top quark properties will contribute to a better understanding of the origin of particle masses and electroweak symmetry breaking, and it might also give hints on the physics that lies beyond the Standard Model. With the large number of top quarks being produced at the LHC \cite{Silva:2012di}, the study of their properties is becoming precision physics. Some observables, such as the total cross section for $t\bar{t}$ production, are expected to be measured with an accuracy of the order of five percent. In order to match these very precise experimental measurements with equally accurate theoretical predictions, next-to-next-to-leading order (NNLO) corrections must be considered.

Although a full NNLO calculation of the $t\bar{t}$ cross section including all required partonic channels is so far missing, a rapidly increasing number of pieces and intermediate results have become available recently \cite{Abelof:2011ap,Anastasiou:2008vd,Baernreuther:2012ws,Bierenbaum:2011gg,Bonciani:2008az,Bonciani:2009nb,Bonciani:2010mn,Czakon:2008zk,Czakon:2011ve,Czakon:2012zr,Kniehl:2008fd,Korner:2008bn}\footnote{An up-to-date review of these intermediate results can be found in \cite{Bonciani:2011zza}.}. Most notably, the inclusive total hadronic $t\bar{t}$ production cross section induced by the all-fermion partonic processes has been computed~\cite{Baernreuther:2012ws,Czakon:2012zr}. 

At NNLO, the calculation of any observable receives three classes of contributions: double real ${\rm d}\sigma^{RR}$, mixed real-virtual, ${\rm d}\sigma^{RV}$, and double virtual contributions ${\rm d}\sigma^{VV}$. For an $m$-jet observable, these contributions contain respectively $(m+2)$, $(m+1)$ and $(m)$ partons in the final state. While the latter contribution is already in an $m$-jet final state configuration, the double real and real virtual classes of partonic channels contribute to the $m$-jet observable at NNLO if the partons present in these channels are theoretically unresolved (soft/collinear) or are experimentally unresolved, i.e clustered to form an $m$-jet final state by a given jet algorithm. In addition, for hadronic observables, mass factorization counterterms which will enter at the $(m+1)$ and $m$-parton level have to be taken into account. The integration of the matrix elements with real radiated particles over the soft and/or collinear regions of phase space yields infrared divergencies. Therefore, in order to evaluate hadronic observables including higher orders, a process independent procedure which enables the extraction and cancellation of those infrared poles amongst the different partonic channels, needs to be applied.

Subtraction methods explicitly constructing analytically integrable infrared subtraction terms which reproduce the behaviour of the full matrix elements in their unresolved limits are well known solutions to this problem \cite{Boughezal:2011jf,Catani:1996vz,Catani:2002hc,Catani:2007vq,Czakon:2010td,Frederix:2008hu,Frixione:1997np,Frixione:2004is,Kilgore:2004ty,Kunszt:1992tn,Nagy:1996bz,Phaf:2001gc,Somogyi:2006cz,Somogyi:2006da,Weinzierl:2003fx}, \cite{Boughezal:2010mc,Daleo:2006xa,Daleo:2009yj,Gehrmann:2011wi,GehrmannDeRidder:2005cm,GehrmannDeRidder:2007jk,GehrmannDeRidder:2011aa,GehrmannDeRidder:2012ja,Glover:2010im}, \cite{Abelof:2011ap,Abelof:2011jv,Abelof:2012rv,Bernreuther:2011jt,GehrmannDeRidder:2009fz}. For QCD observables involving massive fermions, fewer subtraction terms are needed, since the real radiation amplitudes are singular in fewer regions of phase space than their massless counterparts. Indeed for those observables, QCD radiation emitted off a massive leg can only lead to soft singularities. Strict collinear divergencies cannot occur, since they are regulated by the mass of the fermion. In this context, collinear limits are to be replaced by their massive analogues: the quasi-collinear limits \cite{Catani:2002hc,Abelof:2011jv}. In these limits, when integrated over the appropriate phase space, the real radiation matrix element is not divergent but develops terms of the form $\log(m_Q^2/Q^2)$, with $Q$ being the hard scattering scale in the problem under consideration. Since those logarithms are not enhanced in the context of $t \bar{t}$ production at the LHC, they have not been taken into account in the extension of the antenna subtraction method developed in \cite{Abelof:2011ap}. There, subtraction terms capturing only the strict infrared behaviour of the matrix element squared were derived.  Even though, the infrared structure of the matrix elements for processes involving massive fermions is simpler, the kinematics and the integration of the subtraction terms become more involved, given the fact that the finite parton masses introduce a new scale into the problem under consideration.   

At the NLO level, two different subtraction methods have been extended to deal with massive final-state fermions: the dipole formalism~\cite{Catani:2002hc,Phaf:2001gc} and the antenna subtraction method~\cite{Abelof:2011jv,GehrmannDeRidder:2009fz}. The latter has been further extended to NNLO and employed to construct subtraction terms for the double real corrections to heavy quark pair production in \cite{Abelof:2011ap,Abelof:2012rv}. We shall follow this second subtraction framework in this paper.

Based on the universal factorisation properties of QCD colour-ordered amplitudes squared in their infrared limits, antenna subtraction \cite{Boughezal:2010mc,Daleo:2006xa,Daleo:2009yj,Gehrmann:2011wi,GehrmannDeRidder:2005cm,GehrmannDeRidder:2007jk,GehrmannDeRidder:2011aa,GehrmannDeRidder:2012ja,Glover:2010im, Abelof:2011ap,Abelof:2011jv,Abelof:2012rv,Bernreuther:2011jt,GehrmannDeRidder:2009fz} constitutes a framework to construct subtraction terms that approximate the double real and mixed-real virtual corrections in their infrared limits. Within this formalism, the subtraction terms are built as products of antenna functions and reduced matrix elements squared with remapped momenta. 

The antenna functions, which are derived from physical colour-ordered matrix elements, capture all unresolved radiation emitted between two hard radiators and reduce to the well-known universal infrared factors (collinear splitting functions or soft eikonal factors) in the appropriate limits. Depending on whether the two hard radiators are located in the initial or in the final state, three types of antennae are needed: final-final (f-f), initial-final (i-f) and initial-initial (i-i). The initial-final and initial-initial antennae are obtained from their final-final counterparts by crossing one or two final state partons respectively to the initial state. Furthermore, by the flavour of the hard radiators we distinguish between quark-antiquark, quark-gluon and gluon-gluon antennae. While at NLO the subtraction of the unresolved limits of the real corrections only requires tree-level three-parton antennae, at NNLO we need tree-level four-parton antenna functions as well as products of two tree-level three-parton antennae for the double real contributions, and one-loop three-parton antennae for the mixed real-virtual contributions. 

In addition to the antenna functions and the phase space mappings needed for the reduced matrix elements, the method uses a Lorentz invariant factorisation of the full phase space into an antenna phase space, and a reduced phase space with remapped momenta. This factorisation is needed in a different form in each configuration (f-f, i-f, i-i), and it enables the analytic integration of the antenna functions over the antenna phase space. The integrated subtraction terms therefore contain these integrated antennae, and the reduced matrix elements and phase space, both defined in terms of remapped momenta, which are left unintegrated. This integrated subtraction term can then be combined with the virtual contributions and mass factorisation counterterms, achieving thus an analytic cancelation of the infrared poles.

In the massless case, the phase space factorisations have been derived in~\cite{Daleo:2006xa,GehrmannDeRidder:2005cm}. All the massless final-final antennae were integrated in \cite{GehrmannDeRidder:2005cm}, the initial-final ones in \cite{Daleo:2006xa,Daleo:2009yj}, and the initial-initial antennae in~\cite{Boughezal:2010mc,Daleo:2006xa,Gehrmann:2011wi,GehrmannDeRidder:2012ja}. For the antennae involving massive partons, the three-parton tree-level antennae, which are the only ingredients needed at NLO, have been integrated in \cite{Abelof:2011jv,GehrmannDeRidder:2009fz}. At NNLO, so far only one antenna function with massive particles has been integrated \cite{Bernreuther:2011jt}: the final-final four-parton antenna involving a massive and a massless quark-antiquark pair and derived from the physical process $ \gamma^{*} \to Q \bar{Q} q \bar{q}$.

It is the purpose of this paper to evaluate the integrated forms of the new four-parton antenna functions employed in the double real subtraction terms for $t\bar{t}$ production due to the partonic processes $q\bar{q}\to t\bar{t}q'\bar{q}'$ and $gg\to t\bar{t}q\bar{q}$ \cite{Abelof:2011ap,Abelof:2012rv}. These partonic processes constitute the $N_F$ double real contributions to the $t\bar{t}$ hadronic cross section evaluated at NNLO. The three specific antennae have a massless initial-state parton and a massive final-state fermion as their hard radiators, and they are therefore initial-final antennae. 

The paper is organised as follows. In section \ref{sec.antsub}, we briefly recall the infrared structure of double real radiation corrections to hadronic jet observables involving massive final state fermions. We specify in particular how the initial-final four-parton massive antennae enter in the construction of the subtraction terms and present the phase space factorisation that enables the integration of these antennae. In section \ref{sec.antennae} we recall the precise definitions of these specific three initial-final four-parton antennae whose integration constitute the core result of this paper. Section \ref{sec.integration} describes how the integration over the massive initial-final antenna phase space of these four-parton antenna functions is performed. As the expressions of the integrated antennae are too lengthy to be presented here, we shall give the corresponding pole parts in section \ref{sec.integration}, and give the full result in a {\tt Mathematica} file attached separately to this paper. Finally, section 5 contains our conclusions. An appendix containing the unintegrated forms of the initial-final four-parton antennae is included too.

\section{Initial-final antenna subtraction with massive fermions at NNLO}\label{sec.antsub}
At the partonic level, the double real emission contributions to an $m$-jet cross section involving a pair of massive fermions $(Q\bar{Q})$ read 
 \begin{eqnarray}
\lefteqn{{\rm d}\hat\sigma^{RR}_{NNLO}(p_1,p_2)=
{\cal N}_{NNLO}^{RR}\,
\sum_{{m}}{\rm d}\Phi_{m+2}(p_{Q},p_{\bar{Q}},p_{5},\ldots,\,p_{m};
p_1,p_2) }\nonumber \\ && \times 
\frac{1}{S_{{m}}}\,
|{\cal M}_{m+4}(p_{Q},p_{\bar{Q}},p_{5},\ldots,p_{m};p_1,p_2)|^{2}\; 
\JET_{m}^{(m+2)}(p_{Q},p_{\bar{Q}},p_{5},\ldots,p_{m})\hspace{3mm} \\
&\equiv&
{\cal N}_{NNLO}^{RR}\,
\sum_{{m}}{\rm d}\Phi_{m+2}(p_{3}, \ldots, p_{m+4}; p_1,p_2) \nonumber \\ && \times 
\frac{1}{S_{{m}}}\,
|{\cal M}_{m+4}(p_{3},\ldots, p_{m+4},;p_1,p_2)|^{2}\; 
\JET_{m}^{(m+2)}(p_{3},\ldots,p_{m+4}),\hspace{3mm}\label{eq.real}
\end{eqnarray}
where the last line is obtained by relabelling all final state partons. In eq.(\ref{eq.real}), $S_{m}$ is a symmetry factor for identical massless partons in the final state, $|{\cal M}_{m+4}|^2$ denotes a colour-ordered tree-level matrix element squared with $m+2$ final-state partons (two of which are massive) and two are initial-state partons, and $\sum_{m}$ denotes the sum over all the possible colour orderings. The next-to-next-to leading order normalisation factor ${\cal N}^{RR}_{NNLO}$ includes all QCD-independent factors as well as the dependence on the renormalised QCD coupling constant $\alpha_s$.  ${\rm d}\Phi_{m+2}$ denotes the phase space for an $m+2$-parton final state containing $m$ massless and two massive partons with total four-momentum $p_1^{\mu}+p_2^{\mu}$. Finally, the jet function denoted by $\JET_{m}^{(m+2)}$ ensures that out of $m$ massless partons and a $Q\bar{Q}$ pair, an observable with a pair of heavy quark jets in addition to $(m-2)$ jets, is built. 

In principle, only the leading colour pieces of the double real corrections are accounted for in eq.(\ref{eq.real}), since the subleading colour contributions involve in general interferences between sub-amplitudes with different colour orderings. To keep the notation simpler, however, we denote these interference terms also as $|{\cal M}_{m+4}|^2$. 

The NNLO contributions given in eq. (\ref{eq.real}) contain infrared singularities which arise when one or two final state partons are unresolved (soft or collinear). The phase space integration can only be carried out numerically after those singularities have been extracted using subtraction terms that approach the real radiation matrix elements in all their unresolved limits. Depending on the colour connection between the unresolved partons in each sub-amplitude squared in eq.(\ref{eq.real}), the antenna subtraction method distinguishes between the following configurations~\cite{GehrmannDeRidder:2005cm,Glover:2010im}
\begin{itemize}
\item[(a)] One unresolved parton but the experimental observable selects only $m$ jets.
\item[(b)] Two colour-connected unresolved partons (colour-connected).
\item[(c)] Two unresolved partons that are not colour-connected but share a common radiator (almost colour-unconnected).
\item[(d)] Two unresolved partons that are well separated from each other in the colour chain (colour-unconnected).
\item[(e)] Compensation terms for the over-subtraction of large angle soft emission.
\end{itemize}

For each of the configurations listed above, the antenna subtraction terms have a distinct characteristic structure in terms of required antennae functions, which is valid for final-final, initial-final and initial-initial configurations. These characteristic structures, first derived in the massless case in \cite{Daleo:2009yj,GehrmannDeRidder:2005cm,Glover:2010im}, are unaltered by the presence of non-vanishing parton masses \cite{Abelof:2011ap}, although some of the antenna functions needed in this latter case are massive.

With the exceptions of subtraction terms of type $(b)$ and $(e)$, all the remaining configurations are constructed with products of the NLO three-parton tree-level antenna functions. Configuration $(e)$ only arises in partonic processes involving at least three gluons, and its treatment is left to be discussed elsewhere. We shall here focus only on configuration $(b)$ in the presence of massive final-state fermions. In particular, we shall concentrate on those cases where the two unresolved partons are colour-connected to one of the massive final-state particles and to one of the (massless) incoming partons. The form of these specific type of initial-final subtraction terms with massive antennae has been derived in \cite{Abelof:2011ap,Abelof:2012rv} in the context of the evaluation of the double real radiation corrections to top quark pair production at the LHC for different partonic processes. We shall now recall the form of these subtraction terms.

When two unresolved partons $j$ and $k$ are adjacent and colour-connected to two hard radiators labelled $i$ (massless) and $l$ (massive), the initial-final subtraction term ${\rm d}\sigma^{S,b(if)}_{NNLO}$ related to the double real contribution ${\rm d}\sigma^{RR}_{NNLO}$ given in eq.(\ref{eq.real}) reads \cite{Abelof:2011ap}
\beqa
{\rm d}\sigma_{NNLO}^{S,b (if)}&=&  {\cal N}_{NNLO}^{RR}\,\sum_{m}{\rm d}\Phi_{m+2}(p_{3},\ldots,p_{m+4}; p_i,p_2)\frac{1}{S_{{m+2}}} \nonumber \\
&\times& \,\Bigg [ \sum_{jk}\;\left( X^0_{i,jkl}- X^0_{i,jk} X^0_{I,Kl} - X^0_{jkl} X^0_{i,JL} \right)\nonumber \\
&\times&|{\cal M}_{m}(p_{3},\ldots,{p}_{L},\ldots,p_{m+4};x_i
p_i,p_2)|^2\,\JET_{m}^{(m)}(p_{3},\ldots,{p}_{L},\ldots,p_{m+4})\;\Bigg]\;. 
\label{eq.sub2b}
\eeqa
This subtraction term involves: the phase space for the production of $(m+2)$ partons, $\d\Phi_{m+2}$, with two of them being massive, the colour-ordered reduced $(m+2)$-parton amplitude squared $|{\cal M}_{m+2}|^2$ (with two partons less than the original matrix element squared), the jet function $\JET^{(m)}_{m}$, the initial-final four-parton antennae $X_{i,jkl}$ and products of three-parton antennae in final-final and initial-final configurations. The initial-final antenna functions are defined by crossing one massless partons in the corresponding final-final antennae. Since the three and four-parton final-final antennae are denoted by $X_{ijk}$ and $X_{ijkl}$ respectively, the corresponding three and four-parton initial-final antennae are denoted by $X_{i,jk}^0$ and $X_{i,jkl}^0$. 

By construction, the subtraction term in eq.(\ref{eq.sub2b}) contains all colour-connected double unresolved limits of the $(m+4)$-parton colour-ordered matrix elements squared. Those double unresolved limits are captured by the four-parton antennae $X_{i,jkl}$, which are one of the new ingredients appearing in the NNLO double real subtraction terms. The hard radiators are the initial-state parton $i$ and the  (possibly massive) final-state particle $l$, and the unresolved particles are $j$ and $k$. These four-parton antennae, however, are generally also singular in single unresolved limits of $j$ or $k$, where they do not reproduce any physical singularity of the matrix element. In order to ensure that these subtraction terms are only active in the double unresolved regions that they are aimed at, we remove the unphysical single unresolved limits of these four-particle tree-level antennae using appropriate products of two tree-level three-particle antennae, as was done in the massless case in \cite{Daleo:2009yj,Glover:2010im}.

In eq.(\ref{eq.sub2b}), the four-parton antenna functions $X_{i,jkl}$ depend on the original momenta $p_i$, $k_j$, $k_k$ and $k_l$ while the reduced $m$-parton matrix element depend only on the redefined on-shell final-state momenta $p_{3},\ldots,{p}_{L},\ldots p_{m+4}$ and on the rescaled initial-state momenta $x_{i} p_{i}$ and $p_2$. Thus, in order to obtain the integrated form of our subtraction terms, we factorise the full $(m+2)$-particle phase space (with parton $i$ in the initial state) in the following way \cite{Abelof:2011ap}:
\beqa
\d\Phi_{m+2}(p_3,\dots,p_{m+4};p_i,p_2)&=&\d\Phi_{m}(p_3,\dots,p_{L},\dots,p_{m+4};x_i p_i,p_2)\nonumber\\
&\times&\frac{Q^2+m_{jkl}^2}{2\pi}\d\Phi_{3}(p_j,p_k,p_l;p_i,p_2)\frac{\d x_i}{x_i}\,,
\label{eq.psfact}
\eeqa
with $m_{jkl}^2=m_j^2+m_k^2+m_l^2$ and
\beq
x_i=\frac{Q^2+m_{jkl}^2}{2 p_i\cdot q}.
\eeq

Using the phase space factorisation in eq.(\ref{eq.psfact}), the integrated form of the parts of the subtraction terms that involve four-parton antennae are given by
\begin{equation}\label{eq.intsterm}
|{\cal M}_{m+2}|^2\,
J_{m}^{(m)}\; 
{\rm d}\Phi_{m}
\int \frac{Q^2+m_{jkl}^2}{2\pi} {\rm d}\Phi_{3}(p_j,p_k,p_l;p_i,q)\;
X^0_{i,jkl}\frac{{\rm d}x_i}{x_i}.\\
\end{equation}

The integrated form of the four-parton initial-final antennae, denoted by ${\cal X}^0_{i,jkl}$, are thus obtained by integrating the four-parton antennae $X^0_{i,jkl}$ over the $2 \to 3$ particle phase space ${\rm d}\Phi_{3}(p_j,p_k,p_l;p_i,q)$ (with parton $i$ in the initial state), analytically in $d=4-2 \e$ dimensions. More precisely, the integrated four-parton initial-final antenna is defined as: 
\beq\label{eq.aint4}
{\cal X}^0_{i,jkl}(x_i)=\frac{1}{\left[ C(\epsilon)\right]^2} \frac{Q^2+m_{jkl}^2}{2\pi} \int \d\Phi_3(p_j,p_k,p_l;p_i,q) X^0_{i,jkl}\,,
\eeq  
where 
\beq\label{eq.ceps}
C(\epsilon)=(4\pi)^{\epsilon}\frac{e^{-\epsilon \gamma}}{8\pi^2}.
\eeq

The kinematics of the reduced matrix element appearing in eq.(\ref{eq.sub2b}) depends on the masses of the mapped momenta and on the momentum fraction carried by the remapped initial state parton, $x_i$. As will be shown in section \ref{sec.integration}, the integrated antennae also depend on $x_i$, in addition to their natural dependence on the mass of the hard radiator $l$.

\section{Antenna functions for double real radiation}\label{sec.antennae}
In this section we define the three initial-final four-parton massive antenna functions whose integrated form will be given in section \ref{sec.integration}. These antennae have been derived for the first time in \cite{Abelof:2011ap,Abelof:2012rv}, and their explicit expressions will be given in appendix \ref{sec.unintegratedantennae} for completeness. They are present in the subtraction terms needed for the computation of the partonic processes $q\bar{q} \to Q \bar{Q}q'  \bar{q}'$ and $ gg \to Q\bar{Q} q \bar{q}$, which are part of the double real radiation contributions to heavy quark pair production in hadronic collisions. 

In general, an antenna function is characterised by its parton content, and its two radiators. These are the hard partons onto which the antenna collapses to in the unresolved limits. Accordingly, antenna functions are grouped into quark-antiquark, quark-gluon and gluon-gluon antennae and they are all derived from physical matrix elements related, in the final-final case, to the decay of a colourless particle into partons \cite{GehrmannDeRidder:2005aw,GehrmannDeRidder:2005hi}. 

As stated before, while at NLO, only tree-level three-parton antennae are needed to capture the single unresolved behaviour of the real radiation matrix elements, at NNLO, four-parton tree-level antennae and one-loop three-parton antennae are also required. The former are needed to capture the double unresolved behaviour of double real matrix elements while the latter are used to capture the single unresoved behaviour of one-loop matrix elements. Restricting ourselves to the integration of the subtractions terms needed to evaluate double real contributions, we shall not discuss the one-loop antennae in the context of this paper.

In contrast with the massless antenna functions, massive antennae have explicit mass terms and can be of two different natures: flavour-conserving and flavour-violating. The latter are derived from 
partonic processes with a flavour-violating vertex connecting radiators of different flavours, one of them being massless and the other massive. Massive three-parton flavour violating antennae have been derived in~\cite{Abelof:2011jv,Abelof:2012rv}, and one specific quark-antiquark four-parton antenna of this kind has been derived in \cite{Abelof:2011ap}. Since its integrated form will be derived in section \ref{sec.integration}, we shall here recall how it is defined.

When considering the partonic process $q\bar{q} \to Q\bar{Q}q' \bar{q}'$ presented in \cite{Abelof:2011jv}, to account for the singularities that arise when the massless final-state quark-antiquark pair  is unresolved between a massive final state fermion and a massless incoming one, our subtraction terms need a massive initial-final flavour-violating quark-antiquark B-type antenna denoted by $B_4^0(\Q{1},\qb{4},\q{3},\qpi{2})$. In this antenna, the hard radiators are a massive final state quark ($\Q{1}$), and a massless initial state quark ($\qpi{2}$)~\footnote{In the explicit expressions of the antenna functions, we shall label initial state partons with a hat, massless fermions as $q$ and massive fermions as Q.}. 

This antenna is derived from the matrix element of the process $V^* q' \to  Qq\bar{q}$, with a flavour violating vertex joining the colourless virtual particle $V^*$, an initial state massless quark $q'$ and a massive final state quark $Q$. The explicit expression of this antenna together with its infrared limits has been given in \cite{Abelof:2011ap}. It is given in appendix \ref{sec.unintegratedantennae} for completeness. 

The other four-parton antennae whose integrated form is derived in section \ref{sec.integration} are two kinds of quark-gluon antennae. Those are employed in our subtraction terms for the process $gg \to Q\bar{Q} q \bar{q}$ \cite{Abelof:2012rv}, in order to reproduce the double unresolved limits of the double real radiation matrix element in which the final state $q\bar{q}$ pair becomes unresolved between the massive (anti) quark and an initial state gluon. These two quark-gluon antennae are obtained in their final-final forms from the process $\tilde{\chi}\to \tilde{g}gq\bar{q}$, with the massive gluino $\tilde{g}$ playing the role of the massive (anti) quark of mass $m_Q$. The initial-final antennae required here are then obtained by crossing the gluon to the initial state in these final-final ones. The full amplitude for the process $\tilde{\chi}\to \tilde{g}gq\bar{q}$ contains leading and subleading colour pieces~\cite{GehrmannDeRidder:2005aw}. By squaring the leading colour piece, in which the $q\bar{q}$ pair is emitted between the gluino and the gluon in the colour chain, one-kind of quark-gluon antenna, namely the $E_4^0$ antenna, is obtained, while, squaring the subleading colour piece, in which the gluon is emitted between the $q\bar{q}$ pair, the $\wt{E}_4^0$ antenna is obtained. The antenna $E_4^0(\Q{1},\q{3},\qb{4},\gli{2})$ accounts for the infrared limits associated to the emission of an unresolved $q\bar{q}$ pair between a massive (anti) quark and an initial state gluon, while $\wt{E}_4^0(\Q{1},\q{3},\qb{4},\gli{2})$ is used to account for the triple collinear limits that involve a massless $q\bar{q}$ pair and an initial state gluon in those sub-leading colour amplitudes in which the gluon is placed between the quark and the antiquark in the colour chain.

The explicit expressions of these two quark-gluon initial-final antenna functions denoted by $E_4^0(\Q{1},\q{3},\qb{4},\gli{2})$ and $\wt{E}_4^0(\Q{1},\q{3},\qb{4},\gli{2})$ have been derived together with their infrared limits in \cite{Abelof:2012rv} and are recalled in appendix \ref{sec.unintegratedantennae} for completeness.

\section{Integration of initial-final antenna functions at NNLO}\label{sec.integration}
In this section we describe our calculation of the integrated  initial-final four-parton antennae denoted by ${\cal B}^0_{q',Qq\bar{q}}$, ${\cal E}^0_{g,Qq\bar{q}}$ and ${\cal\tilde{E}}^0_{g,Qq\bar{q}}$ whose unintegrated forms, defined in section \ref{sec.antennae} are denoted by $B_4^0(\Q{1},\qb{4},\q{3},\qpi{2})$, $E_4^0(\Q{1},\q{3},\qb{4},\gli{2})$, and $\wt{E}_4^0(\Q{1},\q{3},\qb{4},\gli{2})$ respectively.

\subsection{General structure of the integrated antennae}
The initial-final antennae considered in this paper have a DIS-like $2\rightarrow 3$ kinematics
\beq
q+p_2\rightarrow p_1+p_3+p_4
\eeq
with $p_1^2=m_Q^2$, $p_2^2=p_3^2=p_4^2=0$ and $q^2<0$. It turns out to be convenient to parametrise this kinematics in terms of the following variables
\beq
Q^2=-q^2\, ,\hspace{0.5in}y=1-\frac{Q^2+m_Q^2}{2p_2\cdot q}\, ,\hspace{0.5in} z=\frac{m_Q^2}{E_{cm}^2}
\eeq
where $E_{cm}^2=(p_1+p_3+p_4)^2=(q+p_2)^2$. The more familiar variables $E_{cm}$ and $m_Q$ can be related to $Q^2$, $y$ and $z$ through
\beq\label{eq.relations}
E_{cm}^2=\frac{y}{1-y-z}Q^2\, ,\hspace{0.8in} m_Q^2=\frac{yz}{1-y-z}Q^2.
\eeq
The fact that the $E_{cm}\geq m_Q$ implies that $0\leq z\leq 1$ and $y \geq 0$ and, recalling that $Q^2>0$, it follows from eq.(\ref{eq.relations}) that $y\leq 1-z$. Thus, in terms of our variables $y$ and $z$ the physical region is given by
\beq
0\leq z\leq 1\, , \hspace{0.8in} 0\leq y\leq 1-z.
\eeq

Following eq.(\ref{eq.intsterm}) we integrate the initial-final four-parton antennae over the three particle phase space and express our results in the variables $Q$, $y$ and $z$: ${\cal X}^0_{i,jkl}(Q^2,y,z)$. The integration is carried out following the standard technique of reduction to master integrals using integration-by-parts identities (IBP)~\cite{Chetyrkin:1981qh,Tkachov:1981wb}. 

We start by expressing our phase space integrals as cuts of two-loop four-point functions with two off-shell legs in forward scattering kinematics~\cite{Anastasiou:2002yz}, and then reduce these two-loop integrals using the Laporta algorithm~\cite{Laporta:2001dd} as implemented in {\tt FIRE}~\cite{Smirnov:2008iw}. 

In order to write the phase-space integrals of our four-parton antenna functions as two-loop integrals with cuts, we consider the following propagators:
\beqa
D_1&=&p_1^2-m_Q^2\nonumber\\
D_2&=&p_3^2\nonumber\\
D_3&=&p_4^2\nonumber\\
D_4&=&(p_3+p_4-p_2)^2=(q-p_1)^2\nonumber\\
D_5&=&(p_3+p_4)^2\nonumber\\
D_6&=&(p_3-p_2)^2\nonumber\\
D_7&=&(p_4-p_2)^2\nonumber\\
D_8&=&(p_1+p_3)^2-m_Q^2\nonumber\\
D_9&=&(p_3+p_4-q)^2-m_Q^2=(p_1-p_2)^2-m_Q^2,
\eeqa
where $D_1$, $D_2$ and $D_3$ are cut in the phase space integration. In the reduction procedure we impose momentum conservation $q+p_2=p_1+p_3+p_4$, we set $p_1^2=m_Q^2$, $p_2^2=p_3^2=p_4^2=0$ and $q^2=-Q^2$, and we discard all integrals in which the cut propagators $D_1$, $D_2$ or $D_3$ are not in the denominator.

After carrying out the reduction, we find that the NNLO integrated four-parton antennae ${\cal B}^0_{q',Qq\bar{q}}$, ${\cal E}^0_{g,Qq\bar{q}}$ and ${\cal\tilde{E}}^0_{g,Qq\bar{q}}$ can be expressed in terms of four master integrals
\beqa
I_{[0]}&=&\int \d\Phi_3(p_1,p_2,p_3;p_2,q)\\
I_{[-8]}&=&\int \d\Phi_3(p_1,p_2,p_3;p_2,q)((p_1+p_3)^2-m_Q^2)\\
I_{[4]}&=&\int \d\Phi_3(p_1,p_2,p_3;p_2,q)\frac{1}{(q-p_1)^2}\\
I_{[4,9]}&=&\int \d\Phi_3(p_1,p_2,p_3;p_2,q)\frac{1}{(q-p_1)^2((p_1-p_2)^2-m_Q^2)}.
\eeqa

All these master integrals contain multiplicative factors of the form $y^{-2\e}$ which regulate soft endpoint singularities in initial state convolution integrals. These factors ought to be left as such in the master integrals themselves. All other terms in the master integrals can be expanded in $\e$. Explicitly the master integrals which we denote collectively by $I_{\alpha}(y,z,\e)$ take the form: 
\begin{equation} 
I_{\alpha}(y,z,\e)=y^{m-2\e} R_{\alpha}(y,z,\epsilon).
\end{equation}
The integer $m$ is specific to each master integral and the function $R_{\alpha}(y,z,\epsilon)$ is regular as $y \to 0$ and can be calculated as Laurent series in $\e$.

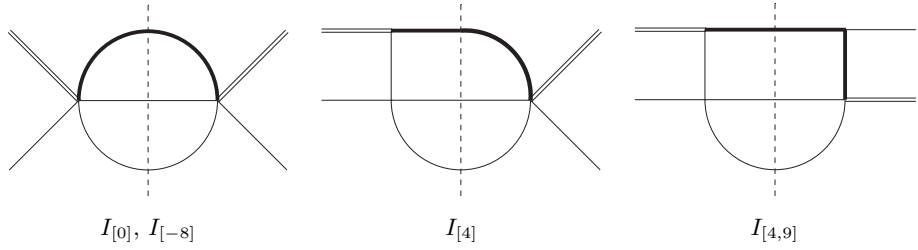
\begin{figure}
  \begin{center}  
  \subfigure[ {$I_{[0]}$, $I_{[-8]}$} ]{
  \resizebox{0.25\linewidth}{!}{
  \begin{picture}(519,352) (157,-111)
    \SetWidth{7.0}
    \SetColor{Black}
    \Arc[clock](416,64)(128,-180,-360)
    \SetWidth{1.0}
    \Arc(416,64)(128,-180,0)
    \Line(288,64)(544,64)
    \Line(288,64)(160,-64)
    \Line[double,sep=6](288,64)(160,192)
    \Line[double,sep=6.5](544,64)(672,192)
    \Line(544,64)(672,-64)
    \Line[dash,dashsize=10](416,240)(416,-112)
  \end{picture}
  } }
  \subfigure[ {$I_{[4]}$} ]{
  \resizebox{0.25\linewidth}{!}{
  \begin{picture}(517,352) (159,-111)
    \SetWidth{8.0}
    \SetColor{Black}
    \Arc[clock](424,72)(120.266,93.814,-3.814)
    \SetWidth{1.0}
    \Arc(416,64)(128,-180,0)
    \Line(288,64)(544,64)
    \Line(288,64)(160,64)
    \Line[double,sep=6](288,192)(160,192)
    \Line[double,sep=6.5](544,64)(672,192)
    \Line(544,64)(672,-64)
    \Line[dash,dashsize=10](416,240)(416,-112)
    \Line(288,192)(288,64)
    \SetWidth{7.0}
    \Line(288,192)(416,192)
  \end{picture}
  } }
  \subfigure[ {$I_{[4,9]}$} ]{
  \resizebox{0.25\linewidth}{!}{
  \begin{picture}(514,352) (159,-111)
    \SetWidth{1.0}
    \SetColor{Black}
    \Arc(416,64)(128,-180,0)
    \Line(288,64)(544,64)
    \Line(288,64)(160,64)
    \Line[double,sep=6](288,192)(160,192)
    \Line[dash,dashsize=10](416,240)(416,-112)
    \Line(288,192)(288,64)
    \SetWidth{7.0}
    \Line(288,192)(544,192)
    \Line(544,192)(544,64)
    \SetWidth{1.0}
    \Line(672,192)(544,192)
    \Line[double,sep=6](672,64)(544,64)
  \end{picture}
  } } 
  \caption{The topologies and mass distributions of the four master integrals encountered in the calculation of the integrated NNLO four-parton antennae ${\cal B}^0_{q',Qq\bar{q}}$, ${\cal E}^0_{g,Qq\bar{q}}$ and ${\cal\tilde{E}}^0_{g,Qq\bar{q}}$. Bold (thin) lines refer to massive (massless)  scalar propagators. The double  line in the external states represents the  off-shell momentum $q$, with $q^2 =  -Q^2$. The cut propagators  are the ones  intersected by the dashed line.}
  \label{topos}
  \end{center}
\end{figure}

The integrated antennae collectively denoted by ${\cal X}(y,z,\e)$ are linear combinations of these master integrals with coefficients containing poles in $\epsilon$ as well as in $y$. After the masters have been inserted into the integrated antennae, these take the form 
\begin{equation}
{\cal X}(y,z,\e) =y^{-1-2\e}  {\cal R}_{{\cal X}}(y,z,\e)
\end{equation}
where ${\cal R_{\cal X}}(y,z,\e)$ is a regular function as $y \to 0$. The $\e$ expansion of the singular factor $ y^{-1-2\e}$ is done in the form of distributions: 
\beq
y^{-1-n\e}=-\frac{\delta(y)}{n\e}+\sum_{m=0}^{\infty}\frac{(-n\e)^m}{m!}{\cal D}_m(y)
\eeq
with
\beq
{\cal D}_m(y)=\left(\frac{\ln ^m(y)}{y} \right)_+\, .
\eeq

It is worth noting that in the functions $R_{\alpha}(y,z,\epsilon)$ and ${\cal R}_{\cal X}(y,z,\e)$, which are regular as $y \to 0$, the limit $z\to 0$, corresponding to the massless limit, cannot be safely taken. In those functions, terms proportional to $\log(z)=\log(m_{Q}^2/E_{cm}^2)$ are present. These terms are expected to be there and correspond to the quasi-collinear kinematical configuration \cite{Abelof:2011jv,Catani:2002hc,GehrmannDeRidder:2009fz}.

To evaluate the integrated antennae, we distinguish two regions depending on the value of $y$: a hard region where $ y\neq 0$ and a soft region where $y=0$. The highest order in $\e$ needed in the expansion of each master integral is determined by the $\e$ and $y$-dependent coefficient that multiplies the integral in the integrated antenna. In the soft region, since the expansion in distribution generates an additional $1/\e$ factor, the function $R_{\alpha}$ is required one order higher in $\e$ than in the hard region.

\subsection{Master integrals}
In this part of the section we present the results concerning the four masters required for the evaluation of the integrated antennae defined above. While $I_{[0]}$ and $I_{[-8]}$ have already been calculated in~\cite{GehrmannDeRidder:2009fz}, $I_{[4]}$ and $I_{[4,9]}$ are presented here for the first time.

\subsubsection{The master integrals $I_{[0]}$ and $I_{[-8]}$}
The master integrals $I_{[0]}$ and $I_{[-8]}$ do not contain any propagator involving the incoming momenta $q, p_2$. Therefore, they are effectively $1\rightarrow 3$ phase space integrals with incoming momentum $q+p_2$ and a massive particle in the final state. These integrals were computed in~\cite{GehrmannDeRidder:2009fz} in the context of the evaluation of final-final three-parton antenna functions with massive final state fermions. They are given by the following all order expressions
\beqa
&&\hspace{-0.2in}I_{[0]}=N_{\epsilon}[C(\e)]^2\left(Q^2+m_Q^2\right)^{1-2\epsilon}\frac{\pi}{2}y^{1-2\epsilon}(1-y)^{-1+2\epsilon}(1-z)^{2-2\epsilon}\nonumber\\
&&\hspace{2.5in}\times\gaussf{1-\epsilon}{2-2\epsilon}{4-4\epsilon}{1-z}\label{eq.I0}\\
&&\hspace{-0.2in}I_{[-8]}=N_{\epsilon}[C(\e)]^2\left(Q^2+m_Q^2\right)^{2-2\epsilon}\frac{\pi}{4}y^{2-2\epsilon}(1-y)^{-2+2\epsilon}(1-z)^{2-2\epsilon}\nonumber\\
&&\hspace{2.5in}\times\gaussf{1-\epsilon}{2-2\epsilon}{5-4\epsilon}{1-z}, \label{eq.I-8}
\eeqa
with
\beq
N_{\epsilon}=e^{2\epsilon\gamma_E}\frac{\Gamma(1-\epsilon)^2}{\Gamma(4-4\epsilon)}.
\eeq
The hypergeometric functions in eqs.(\ref{eq.I0}) and (\ref{eq.I-8}) can be expanded with the {\tt Mathema \newline tica} package {\tt HypExp}~\cite{Huber:2005yg}, yielding ordinary harmonic polylogarithms (HPLs) \cite{Remiddi:1999ew} in the variable $z$. In the computation of the integrated antennae of this paper, we find that these two integrals are needed in the hard regions up to ${\cal O}(\e^2)$ while in the soft region, they are needed up to ${\cal O}(\e^3)$. At this order (${\cal O}(\e^3)$), the harmonic polylogarithms (HPLs) in the variable $z$ are needed up to weight 3.

\subsubsection{The master integrals $I_{[4]}$ and $I_{[4,9]}$}
To evaluate the two remaining master integrals we use the method of differential equations~\cite{Caffo:1998du,Caffo:1998yd,Gehrmann:1999as,Kotikov:1990kg,Kotikov:1991hm,Kotikov:1991pm,Remiddi:1997ny}. In order to obtain those we proceed as follows: We start by noting that
\beqa
q_{\mu}\frac{\partial\,\,}{\partial q_{\mu}}&=&2Q^2\frac{\partial\,\,}{\partial Q^2}-\frac{1 - y - z - y z}{1-z}\frac{\partial\,}{\partial x}+\frac{1 - 2 y - z}{y}\frac{\partial\,}{\partial z}\nonumber\\
p_{2\mu}\frac{\partial\,\,}{\partial p_{2\mu}}&=&(1-y)\frac{\partial\,}{\partial y}-\frac{z(1-z)}{y}\frac{\partial\,}{\partial z}\nonumber\\
m_Q^2\frac{\partial\,\,}{\partial m_Q^2}&=&-\frac{yz}{1-z}\frac{\partial\,\,}{\partial y}+z\frac{\partial\,}{\partial z},
\eeqa
and invert this system to get
\beqa
Q^2\frac{\partial\,\,}{\partial Q^2}&=&\frac{1}{2}\,q_{\mu}\frac{\partial\,}{\partial q_{\mu}}+\frac{1}{2}\,p_{2\mu}\frac{\partial\,}{\partial p_{2\mu}}+\frac{Q^2 yz}{1-y-z}\,\frac{\partial\,}{\partial m_Q^2}\nonumber\\
\frac{\partial\,}{\partial y}&=&\frac{1}{1-y-z}\,p_{2\mu}\frac{\partial\,}{\partial p_{2\mu}}+\frac{Q^2 z(1-z)}{(1-y-z)^2}\,\frac{\partial\,}{\partial m_Q^2}\nonumber\\
\frac{\partial\,}{\partial z}&=&\frac{y}{(1-z)(1-y-z)}\,p_{2\mu}\frac{\partial\,}{\partial p_{2\mu}}+\frac{Q^2 y(1-y)}{(1-y-z)^2}\,\frac{\partial\,}{\partial m_Q^2}.\label{eq.diffeq}
\eeqa
We then apply the differential operators in eq.(\ref{eq.diffeq}) to the integrals $I_{[4]}$ and $I_{[4,9]}$ and reduce the resulting integrals on the right hand side to master integrals. We thus obtain a set of first order partial differential equations for the master integrals which can be readily solved with standard techniques. The solution is found order by order in a Laurent expansion in $\e$. This expansion involves harmonic polylogarithmic functions of one variable (HPL's), which can be functions of either $y$ or $z$, or generalised harmonic polylogarithmic functions (GHPL's) \cite{Gehrmann:2000xj,Gehrmann:2001jv} of variable $z$ and arguments which can are taken from the list  $(0,1,1-y)$. Products of these two type of functions are also found.

The master integral $I_{[4]}$ is finite. For the integrated antennae required  in this paper, it is needed in the hard region ($y \neq 0$) up to $\order{\e^2}$ while in the soft region it is needed up to $\order{\e^3}$. At this order ($\order{\e^3}$) it will involve polylogarithmic functions, whose overall weight can go up to 4, as the $\order{\e^0}$ involves polylogarithmic functions of weight 1. This master integral can, however, be calculated in both regions up to order $\e^3$. In the hard region $(y\neq 0)$ it is given by
\beqa
I_{[4]}&=&[C(\e)]^2\left(Q^2+m_Q^2\right)^{-2\e}\frac{\pi}{2}y^{-2\e}(1-y-z)^{-1}\bigg\{-(1-y) (1-z) H(1;y)-y z H(0;z)\nonumber\\
&&+\e\bigg[(1-y)(1-z) \bigg(-2 H(1;y) G(1-y;z)-2 G(1-y,0;z)-2 H(1;y) H(1;z)\nonumber\\
&& -2 H(0,1;y)+3 H(1,1;y)\bigg)+ y z \bigg(2 H(1;y)H(0;z)-5 H(0;z)+H(0,0;z)\nonumber\\
&&-5 H(1;y)-2 H(0,1;z)+\frac{2 \pi ^2}{3}\bigg)-2 (1-y-z) H(1,0;z)\bigg]\nonumber\\
&&+\e^2\bigg[(1-y)(1-z) \bigg(-10 H(1;y) G(1-y;z)-4 H(0,1;y) G(1-y;z)\nonumber\\
&&+6 H(1,1;y) G(1-y;z)+4 H(1;y) G(1,1-y;z)+4 H(1;y) G(1-y,0;z)\nonumber\\
&&+4H(1;y) G(1-y,1;z)-4 H(1;y) G(1-y,1-y;z)+\frac{4}{3} \pi ^2 G(1-y;z)\nonumber\\
&&-10 G(1-y,0;z)+4 G(1,1-y,0;z)+2 G(1-y,0,0;z)+4G(1-y,0,1;z)\nonumber\\
&&-4 G(1-y,1-y,0;z)-10 H(1;y) H(1;z)-4 H(0,1;y) H(1;z)\nonumber\\
&&+6 H(1,1;y) H(1;z)-4 H(1;y) H(1,1;z)-19 H(1;y)-10H(0,1;y)\nonumber\\
&&+15 H(1,1;y)-4 H(0,0,1;y)+6 H(0,1,1;y)+2 H(1,0,1;y)-7 H(1,1,1;y)\bigg)\nonumber\\
&&+y z \bigg(10 H(1;y) H(0;z)-4H(1,1;y) H(0;z)-2 H(1;y) H(0,0;z)\nonumber\\
&&+4 H(1;y) H(0,1;z)-\frac{1}{2} \pi ^2 H(1;y)+\frac{1}{6} \left(\pi ^2-114\right) H(0;z)\nonumber\\
&&+5H(0,0;z)-10 H(0,1;z)-H(0,0,0;z)+2 H(0,0,1;z)-4 H(0,1,0;z)\nonumber\\
&&-4 H(0,1,1;z)-4 H(1,1,0;z)+\frac{2}{3} \left(12 \zeta_3+5 \pi^2\right)\bigg)\nonumber\\
&&+(1-y-z) \bigg(4 H(1;y) H(1,0;z)+\frac{4}{3} \pi ^2 H(1;z)-10 H(1,0;z)\nonumber\\
&&+2 H(1,0,0;z)-4 H(1,0,1;z)-\frac{5}{6}\pi ^2 z H(1;y)\bigg)\bigg]\nonumber\\
&&+\e^3\bigg[(1-y)(1-z) \bigg(-\left(38+\pi ^2\right) H(1;y) G(1-y;z)\nonumber\\
&&-20 H(0,1;y) G(1-y;z)+30 H(1,1;y)G(1-y;z)-8 H(0,0,1;y) G(1-y;z)\nonumber\\
&&+12 H(0,1,1;y) G(1-y;z)+4 H(1,0,1;y) G(1-y;z)-14 H(1,1,1;y) G(1-y;z)\nonumber\\
&&+\frac{4}{3} \left(5 \pi^2+12 \zeta_3\right) G(1-y;z)\nonumber\\
&&-\frac{8}{3} \pi ^2 G(1,1-y;z)+\frac{1}{3} \left(-114+\pi ^2\right) G(1-y,0;z)\nonumber\\
&&+\frac{8}{3} \pi ^2G(1-y,1-y;z)+20 G(1,1-y,0;z)+10 G(1-y,0,0;z)\nonumber\\
&&+20 G(1-y,0,1;z)-20 G(1-y,1-y,0;z)-8 G(1,1,1-y,0;z)\nonumber\\
&&-4 G(1,1-y,0,0;z)-8G(1,1-y,0,1;z)+8 G(1,1-y,1-y,0;z)\nonumber\\
&&-2 G(1-y,0,0,0;z)-4 G(1-y,0,0,1;z)+8 G(1-y,0,1,0;z)\nonumber\\
&&-8 G(1-y,0,1,1;z)-8 G(1-y,1,1,0;z)+8G(1-y,1,1-y,0;z)\nonumber\\
&&+4 G(1-y,1-y,0,0;z)+8 G(1-y,1-y,0,1;z)-8 G(1-y,1-y,1-y,0;z)\nonumber\\
&&+20 G(1,1-y;z) H(1;y)+20 G(1-y,0;z) H(1;y)+20G(1-y,1;z) H(1;y)\nonumber\\
&&-20 G(1-y,1-y;z) H(1;y)-8 G(1,1,1-y;z) H(1;y)\nonumber\\
&&-8 G(1,1-y,0;z) H(1;y)-8 G(1,1-y,1;z) H(1;y)+8G(1,1-y,1-y;z) H(1;y)\nonumber\\
&&-4 G(1-y,0,0;z) H(1;y)-8 G(1-y,0,1;z) H(1;y)-8 G(1-y,1,1;z) H(1;y)\nonumber\\
&&+8 G(1-y,1,1-y;z) H(1;y)+8G(1-y,1-y,0;z) H(1;y)\nonumber\\
&&+8 G(1-y,1-y,1;z) H(1;y)-8 G(1-y,1-y,1-y;z) H(1;y)-65 H(1;y)\nonumber\\
&&-38 H(1;y) H(1;z)+8 G(1,1-y;z)H(0,1;y)+8 G(1-y,1;z) H(0,1;y)\nonumber\\
&&-8 G(1-y,1-y;z) H(0,1;y)-20 H(1;z) H(0,1;y)\nonumber\\
&&-\frac{1}{3} \left(114-5 \pi ^2\right) H(0,1;y)-12G(1,1-y;z) H(1,1;y)\nonumber\\
&&-8 G(1-y,0;z) H(1,1;y)-12 G(1-y,1;z) H(1,1;y)+12 G(1-y,1-y;z) H(1,1;y)\nonumber\\
&&+30 H(1;z) H(1,1;y)+57H(1,1;y)-20 H(1;y) H(1,1;z)-8 H(0,1;y) H(1,1;z)\nonumber\\
&&+12 H(1,1;y) H(1,1;z)-8 H(1;z) H(0,0,1;y)-20 H(0,0,1;y)\nonumber\\
&&+12 H(1;z)H(0,1,1;y)+30 H(0,1,1;y)+4 H(1;z) H(1,0,1;y)\nonumber\\
&&+10 H(1,0,1;y)-14 H(1;z) H(1,1,1;y)-35 H(1,1,1;y)-8 H(1;y) H(1,1,1;z)\nonumber\\
&&-8H(0,0,0,1;y)+12 H(0,0,1,1;y)+4 H(0,1,0,1;y)-14 H(0,1,1,1;y)\nonumber\\
&&+4 H(1,0,0,1;y)-6 H(1,0,1,1;y)-2 H(1,1,0,1;y)+15H(1,1,1,1;y)\bigg)\nonumber\\
&&+y z \bigg(H(0,0,0,0;z)+4H(0,1,1,0;z)+4 H(1,1,0,0;z)\nonumber\\
&&-\frac{1}{3} \left(-114+\pi ^2\right) H(0;z) H(1;y)-\frac{5}{2} \pi ^2 H(1;y)\nonumber\\
&&+\frac{5}{3} \pi ^2 H(1;y)H(1;z)-10 H(1;y) H(0,0;z)\nonumber\\
&&+\frac{1}{6} \left(114-\pi ^2\right) H(0,0;z)+20 H(1;y) H(0,1;z)\nonumber\\
&&-\left(38-\frac{13 \pi ^2}{3}\right)H(0,1;z)-20 H(0;z) H(1,1;y)\nonumber\\
&&+4 H(0,0;z) H(1,1;y)-8 H(0,1;z) H(1,1;y)+\frac{8}{3} \pi ^2 H(1,1;z)\nonumber\\
&&+2 H(1;y) H(0,0,0;z)-5 H(0,0,0;z)-4 H(1;y) H(0,0,1;z)\nonumber\\
&&+10 H(0,0,1;z)+8 H(1;y) H(0,1,0;z)-20 H(0,1,0;z)+8 H(1;y) H(0,1,1;z)\nonumber\\
&&-20 H(0,1,1;z)+8H(1;y) H(1,1,0;z)-20 H(1,1,0;z)+8 H(0;z) H(1,1,1;y)\nonumber\\
&&-2 H(0,0,0,1;z)+4 H(0,0,1,0;z)+4 H(0,0,1,1;z)+4 H(0,1,0,0;z)\nonumber\\
&&-8H(0,1,0,1;z)-8 H(0,1,1,1;z)-8 H(1,1,0,1;z)+\frac{1}{9} \left(114 \pi ^2-\pi ^4+360 \zeta_3\right)\nonumber\\
&&+\frac{1}{6} H(0;z) \left(-390+5 \pi ^2+16\zeta_3\right)\nonumber\\
&&-\frac{1}{6}\pi^2 H(1,1,y)-\frac{16}{3}\zeta_3 \bigg)\nonumber\\
&&+(1-y-z) \bigg(-8 H(1,0,1,0;z)+8 H(1,1,1,0;z)-\pi ^2 H(1;y) H(1;z)\nonumber\\
&&+20 H(1;y) H(1,0;z)-\frac{1}{3}\left(114-\pi ^2\right) H(1,0;z)\nonumber\\
&&-8 H(1,0;z) H(1,1;y)-4 H(1;y) H(1,0,0;z)+10 H(1,0,0;z)+8 H(1;y) H(1,0,1;z)\nonumber\\
&&-20 H(1,0,1;z)-2H(1,0,0,0;z)+4 H(1,0,0,1;z)-8 H(1,0,1,1;z)\nonumber\\
&&+\frac{4}{3} H(1;z) \left(5 \pi ^2+12 \zeta_3\right)-\frac{5}{2}\pi^2 H(1,1;y)\nonumber\\
&&+\left(\frac{25}{6}\pi^2+\frac{32}{3}\zeta_3 \right) H(1;y)\bigg)\bigg]\nonumber\\
&&+{\cal O}(\e^4)\bigg\}.\label{eq.I4}
\eeqa
The expression of $I_{[4]}$ in the soft region is simply obtained by setting $y=0$ in eq.(\ref{eq.I4}). 

The master integral $I_{[4,9]}$ appears in the reduction of the integrated antennae into masters. It is finite, and a priori only its soft limit is needed up to order $\order{\e^0}$. Up to this order, its expression in the hard region reads
\beqa
I_{[4,9]}&=&[C(\e)]^2\left(Q^2+m_Q^2\right)^{-2\e}\frac{\pi}{2}y^{-2\e}(1-y)\bigg[H(1;y)H(0;z)\nonumber\\
&&-H(1;y) G(1-y;z)-G(1-y,0;z)+H(0,1;y)+H(1,1;y)-H(1,0;z)+{\cal O}(\e)\bigg].\nonumber\\
\eeqa
With the soft limit of this expression being zero, it turns out that this master integral does not contribute in the result obtained for the integrated antennae derived in this paper.

\subsection{Integrated antennae}
The three integrated antennae ${\cal B}^0_{q',Qq\bar{q}}$, ${\cal E}^0_{g,Qq\bar{q}}$ and ${\cal\tilde{E}}^0_{g,Qq\bar{q}}$ have a common multiplicative factor given by $(Q^2 +m_Q^2)^{-2\e}$. The full expression of these integrated antennae are too lenghty to be presented here. A separate {\tt Mathematica} file containing them is attached with the arXiv submission of this paper. In this section we shall present the complete pole part of these integrated initial-final four-parton antennae. 

The leading pole parts of ${\cal B}^0_{q',Qq\bar{q}}$ and ${\cal E}^0_{g,Qq\bar{q}}$ are obtained in the kinematical configuration where the final-state $q \bar{q}$ pair becomes soft and simultaneously collinear to an initial-state parton. Those leading poles parts are proportional to $1/\e^3\delta(y)$. 

Having no soft $q\bar{q}$ singularities, the integrated antenna ${\cal\tilde{E}}^0_{g,Qq\bar{q}}$ is regular in the limit $y\to 0$ and it does not contain any distributions. Its leading pole piece is proportional to $1/\e^2$.

To make our expressions more concise we define the following variable
\beq
\lambda=\sqrt{\frac{y\,z}{1-y-z}}
\eeq
which is equal to $m_{Q}/Q$.

Using this notation, the pole parts of the integrated antennae are:
\beqa
\lefteqn{ {\cal B}^0_{q',Qq\bar{q}}=\left( Q^2+m_Q^2\right)^{-2\epsilon}}\nonumber\\
&&\times\bigg\{ -\frac{1}{\epsilon^3}\frac{1}{12}\delta(y) \nonumber\\
&& +\frac{1}{\epsilon^2}\left[ \delta(y)\left( -\frac{19}{72}-\frac{z}{12}+\frac{z^2}{24} \right) + \frac{1}{6}D_0(y) + \frac{1}{6}\delta(y) G(1;z)-\frac{1}{6}+\frac{y}{12}     \right]    \nonumber\\
&& +\frac{1}{\epsilon} \bigg[\delta(y)\left( -\frac{373}{432}+\frac{5\pi^2}{72}-\frac{19z}{72}+\frac{25z^2}{144} \right) +D_0(y)\left( \frac{19}{36}+\frac{z}{6}-\frac{z^2}{12} \right)\nonumber\\
&&\hspace{0.3in} -\frac{1}{3}D_1(y) -\frac{11}{18}+\frac{5y}{36}+\frac{z}{1-z}\delta(y)\left( \frac{1}{6}+\frac{z}{8}-\frac{z^2}{24} \right)H(0;z)\nonumber\\
&&\hspace{0.3in} -\delta(y)\left(\frac{19}{36}+\frac{z}{6}-\frac{z^2}{12}\right)H(1;z)  -\frac{1}{6}\delta(y)H(1,0;z)-\frac{1}{3}\delta(y)H(1,1;z)\nonumber\\
&&\hspace{0.3in}  +\frac{1}{3}D_0(y)H(1;z)+\left(\frac{1}{3}-\frac{y}{6}  \right)H(0;y)-\left( \frac{1}{2y}-\frac{1}{2}+\frac{y}{4}\right)H(1;y)-\left( \frac{1}{3}-\frac{y}{6}\right) H(1;z)   \bigg]\nonumber\\
&& + {\cal O}(\epsilon^0)\bigg\},
\eeqa

\beqa
\lefteqn{ {\cal E}^0_{g,Qq\bar{q}}=\left( Q^2+m_Q^2\right)^{-2\epsilon}}\nonumber\\
&&\times\bigg\{ -\frac{1}{\epsilon^3}\frac{1}{12}\delta(y) \nonumber\\
&& +\frac{1}{\epsilon^2}\bigg[ \delta(y)\left( -\frac{19}{72}-\frac{z}{12}+\frac{z^2}{24} \right)+\frac{1}{6}D_0(y)-\frac{1}{6}\delta(y)H(1;z)\nonumber\\
&&\hspace{0.3in} +\frac{1}{1-y}\left( -\frac{1}{6}+\frac{9y}{8}-\frac{23y^2}{24}+\frac{y^3}{3} \right)-\left(\frac{1}{2}-\frac{y}{4} \right)H(1;y)\bigg]\nonumber\\
&&+\frac{1}{\epsilon}\bigg[ \delta(y)\bigg( -\frac{373}{432}+\frac{5\pi^2}{72}-\frac{19z}{72}+\frac{25z^2}{144}\bigg)+D_0(y)\bigg( \frac{19}{36}+\frac{z}{6}-\frac{z^2}{12}\bigg)-\frac{1}{3}D_1(y)\nonumber\\
&&\hspace{0.3in}+\frac{1}{1-y}\bigg( \frac{1}{18}-\frac{\lambda}{3}+\frac{7y}{2}-\frac{29y^2}{9}-\frac{\lambda y^2}{2}+\frac{4y^3}{3}-\frac{z}{6}-\frac{z^2}{12}\bigg)+\frac{1}{(1-y)(1-z)}\bigg( \frac{y}{3}+\frac{\lambda y}{3}\nonumber\\
&&\hspace{0.3in}-\frac{y^2}{2}+\lambda y^2 -\frac{2y^3}{3}+\frac{\lambda y^3}{2}   \bigg)+\frac{y^3}{(1-y)(1-z)^2}\bigg( \frac{2}{3}-\lambda\bigg)-\delta(y)\bigg( \frac{1}{4}-\frac{1}{4(1-z)}+\frac{z}{12}\nonumber\\
&&\hspace{0.3in}-\frac{z^2}{24}\bigg)H(0;z)-\delta(y)\bigg( \frac{19}{36}+\frac{z}{6}-\frac{z^2}{12}\bigg)H(1;z)-\frac{1}{6}\delta(y)H(1,0;z)-\frac{1}{3}\delta(y)H(1,1;z)\nonumber\\
&&\hspace{0.3in}+\frac{1}{3}D_0(y)H(1;z)+\bigg(1-\frac{2}{3(1-y)}-\frac{5y}{4}+\frac{2y^2}{3}\bigg)H(0;y)+\bigg( \frac{y(1-y)}{2(1-y-z)}\nonumber\\
&&\hspace{0.3in}+\frac{1}{6}-\frac{\lambda}{2}-\frac{2}{3(1-y)}-\frac{1}{3y}-\frac{11y}{24}+\frac{2y^2}{3}   \bigg) H(1;y)\nonumber\\
&&\hspace{0.3in}+\bigg( -\frac{2}{3}-\frac{1}{2}+\frac{1-5\lambda}{6(1-y)}+\frac{3y}{2}-\frac{2y^2}{3}-\frac{y(1-y)}{2(1-y-z)}+\frac{1}{1-z}\bigg( -\frac{3y}{2}+\frac{11\lambda}{6}\nonumber\\
&&\hspace{0.3in}+\lambda y +\frac{2-13\lambda}{6(1-y)}\bigg)-\frac{1}{(1-z)^2}\bigg(-\frac{7}{6}-\frac{3y}{2}+\frac{7\lambda}{3} +2\lambda y -y^2+\lambda y^2 +\frac{7(1-2\lambda)}{6(1-y)}   \bigg)\nonumber\\
&&\hspace{0.3in} +\frac{(2-3\lambda)y^3}{2(1-y)(1-z)^3} \bigg)H(0;z)+\bigg( -1+\frac{2}{3(1-y)}+\frac{5y}{4}-\frac{2y^2}{3} \bigg)H(1;z)\nonumber\\
&&\hspace{0.3in}-\bigg(1-\frac{y}{2}\bigg)H(1;y)H(1;z)-\bigg(1-\frac{y}{2}\bigg)H(1;y)G(1-y;z)+\bigg(1-\frac{y}{2}\bigg)H(1,0;y)\nonumber\\
&&\hspace{0.3in}+\bigg( \frac{3}{2}-\frac{3y}{4}\bigg)H(1,1;y)-\bigg(1-\frac{y}{2}\bigg)H(1,0;z)-\bigg(1-\frac{y}{2}\bigg)G(1-y,0;z)
\bigg]\nonumber\\
&&+{\cal O}(\epsilon^0)\bigg\},
\eeqa

\beqa
\lefteqn{ {\cal \tilde{E}}^0_{g,Qq\bar{q}}=\left( Q^2+m_Q^2\right)^{-2\epsilon}}\nonumber\\
&&\times\bigg\{\frac{1}{\epsilon^2}\bigg[-\frac{1}{3}+\frac{1}{3(1-y)}+\frac{11y}{12}-\frac{y^2}{3}-\bigg( 1-\frac{y}{2}\bigg)H(1;y)     \bigg]\nonumber\\
&&+\frac{1}{\epsilon}\bigg[-\frac{20}{9}+\frac{16y}{9}+\lambda y-\frac{13y^2}{18}+\lambda+\frac{1}{1-y}\bigg(\frac{20}{9}-\lambda\bigg)\nonumber\\
&&\hspace{0.3in}+\frac{y^2}{(1-y)(1-z)}\bigg( -1+2\lambda-\frac{4y}{3}+\lambda y  \bigg)+\frac{y^3}{(1-y)(1-z)^2}\bigg(\frac{4}{3}-2\lambda\bigg)\nonumber\\
&&\hspace{0.3in}+\bigg(\frac{2}{3}-\frac{11y}{6}+\frac{2y^2}{3}-\frac{2}{3(1-y)}\bigg)H(0;y)+\bigg(-\frac{1}{3}-\lambda-\frac{7y}{12}+\frac{2y^2}{3}-\frac{2}{3(1-y)}\nonumber\\
&&\hspace{0.3in} -\frac{y-y^2}{1-y-z}   \bigg)H(1;y)+\bigg( \frac{y}{1-y}\bigg( 2+\lambda-3y+\frac{2y}{3} \bigg) -\frac{y(1-y)}{1-y-z}\nonumber\\
&&\hspace{0.3in}-\frac{y}{(1-y)(1-z)}\left( 1+\lambda-3y +2\lambda y \right) +\frac{y^2}{(1-y)(1-z)^2}\left( -1+2\lambda-2y+2\lambda y\right) \nonumber\\
&&\hspace{0.3in}+ \frac{y^3}{(1-y)(1-z)^3}\bigg( \frac{4}{3}-\lambda \bigg)  \bigg)H(0;z)+\frac{y}{(1-y)}\bigg( \frac{15}{6}-\frac{15y}{6}+\frac{2y^2}{3}\bigg)H(1;z)\nonumber\\
&&\hspace{0.3in}+(2-y)H(1;y)H(1;y)-(2-y)H(1;y)G(1-y;z)+\bigg(3-\frac{3y}{2}\bigg)H(1,1;y)\nonumber\\
&&\hspace{0.3in}+(2-y)H(1,0;y)-(2-y)H(1,0;z)-(2-y)G(1-y,0;z)
\bigg]\nonumber\\
&&+{\cal O}(\epsilon^0)\bigg\}.
\eeqa

\section{Conclusions}
We have derived the integrated forms of the massive initial-final tree-level four-parton antennae appearing in the subtraction terms required to compute the partonic processes $ q \bar{q} \to t \bar{t} q' \bar{q}'$ and $gg \to t \bar{t}q\bar{q}$ contributing to the double real corrections to $t \bar{t}$ hadronic production. We found that those integrated antennae can be written as combinations of three master integrals from which one was unknown. This master integral is derived using differential equation methods and presented for the first time in this paper. These integrated antennae, related to the terms proportional to the colour factor $N_{F}$ in top quark pair production at NNLO represent the core result of this paper. With the integrated forms of these antenna functions now available, the integrated subtraction terms containing these antennae can be easily computed and combined with the double-virtual, one-parton integrated real-virtual and mass factorisation counterterms participating in the $m$-parton final state channel of an $m$-jet observable evaluated at NNLO. These integrated subtraction terms will contribute in a non-trivial way to the cancellation of the explicit poles present in this $m$-parton channel. As such, the results presented in this paper are essential for the application of the extended antenna formalism to compute hadronic observables with massive fermions at the NNLO level. These results, furthermore enable the computation of the $N_{F}$ contributions to the hadronic $t \bar{t}$ cross section evaluated at NNLO.

\section{Acknowledgments}
We would like to thank Werner Bernreuther for comments on our manuscript. Oliver Dekkers would like to thank the institute for theoretical physics at ETH Zurich, where most of this research project has been carried out, for its kind hospitality. This research was supported by the Swiss National Science Foundation (SNF) under contract PP00P2-139192 and in part by the European Commission through the 'LHCPhenoNet' Initial Training Network PITN-GA-2010-264564', which are hereby acknowledged. Furthermore, the work of Oliver Dekkers was supported by the Deutsche Forschungsgemeinschaft (DFG), SFB/TR9.

\appendix
\section{Four-parton antennae}\label{sec.unintegratedantennae}
In this section we list the expressions of the unintegrated initial-final four-parton massive antennae defined in section \ref{sec.antennae}.

The flavour-violating initial-final four-parton B-type antenna is given by:
\beqa
B_4^0(\Q{1},\qb{4},\q{3},\qpi{2})&=&-\frac{1}{\left(Q^2+m_Q^2\right)}\bigg\{ -\frac{1}{s_{34}s_{134}^2}\left( s_{12}s_{13}+s_{12}s_{14}+s_{13}s_{23}+s_{14}s_{24}\right)\nonumber\\
&&+\frac{1}{s_{34}s_{234}^2}\left( s_{12}s_{23}+s_{12}s_{24}-s_{13}s_{23}-s_{14}s_{24}\right)\nonumber\\
&&-\frac{1}{s_{34}^2 s_{134}^2}\left( 2s_{12}s_{13}s_{14}+s_{13}s_{14}s_{24}+s_{13}s_{14}s_{23}-s_{13}^2 s_{24}-s_{14}^2 s_{23}\right)\nonumber\\
&&+\frac{1}{s_{34}^2 s_{234}^2}\left( -2s_{12}s_{23}s_{24}+s_{13}s_{23}s_{24}+s_{14}s_{23}s_{24}-s_{13} s_{24}^2-s_{14} s_{23}^2\right)\nonumber\\
&&+\frac{1}{s_{34}s_{134}s_{234}}\left( 2s_{12}^2+s_{12}s_{23}+s_{12}s_{24}-s_{12}s_{13}-s_{12}s_{14}\right)\nonumber\\
&&+\frac{1}{s_{34}^2s_{134}s_{234}}\big( -s_{13}s_{24}^2-s_{14}s_{23}^2+s_{13}^2 s_{24}+s_{14}^2 s_{23}\nonumber\\
&& -s_{13}s_{14}s_{23}-s_{13}s_{14}s_{24}+s_{13}s_{23}s_{24}+s_{14}s_{23}s_{24} \nonumber\\
&& -2s_{12}s_{13}s_{24}-2s_{12}s_{14}s_{23}\big)-\frac{2s_{12}}{s_{134}s_{234}}\nonumber\\
&& +2m_Q^2\bigg( \frac{1}{s_{134}^2}+\frac{1}{s_{134}s_{234}}-\frac{1}{s_{34}s_{134}}+\frac{s_{12}-s_{234}}{s_{34}s_{134}^2} \bigg)\bigg\}+\order{\epsilon},\label{eq.B04fl}
\eeqa
where $Q^2=-(p_1-p_2+p_3+p_4)^2$, $s_{134}=s_{13}+s_{14}+s_{34}$, and $s_{234}=-s_{23}-s_{24}+s_{34}$. In this case the normalisation is
\beq
\left| \cm_2(\gamma^*q\rightarrow Q)\right|^2=4(1-\epsilon)(Q^2+m_Q^2).
\eeq\\

The initial-final quark-gluon four-parton E-type antenna is given by:
\beqa 
E_4^0(\Q{1},\q{3},\qb{4},\gli{2})&=&\frac{1}{\left(Q^2+m_Q^2\right)^2}\bigg\{
\frac{s_{13} s_{14}} {s_{12} s_{234}}  + \frac{s_{14}}{ s_{12} s_{34} s_{24}}  \left[- s_{13} s_{23}+ s_{13}^2 + s_{14}^2 \right]\nonumber \\ 
&& + \frac{s_{14}}{ s_{12} s_{34} s_{134}}  \left[ 2 s_{14} s_{23}+ 2 s_{14} s_{24}+ s_{23}^2- s_{24}^2\right]\nonumber \\ 
&&+ \frac{s_{14}}{ s_{12} s_{24} s_{234}}  \left[-s_{13} s_{23}+ s_{13}^2+ s_{14}^2\right]+ \frac{s_{14}}{ s_{12} s_{24}}  \left[2 s_{13}+ 2 s_{14}-s_{23}\right]\nonumber \\ 
&& + \frac{4 s_{14} s_{12}^2s_{23}}{ s_{34}^2  s_{134} s_{234} }  + \frac{s_{14}}{ s_{34}^2 s_{134}}  \Big[ +4  s_{12}s_{14} + 4 s_{14} s_{23}+ 4 s_{14}s_{24}\nonumber \\ 
&&+ 8 s_{12} s_{24}+ 4 s_{12}^2+ 4 s_{23} s_{24}+ 4 s_{24}^2\Big]\nonumber \\ 
&&- \frac{s_{14}}{ s_{34} s_{24} s_{134}}  \left[ 2 s_{12}s_{14} + s_{14} s_{23}- 2 s_{14}^2- s_{12} s_{23}- s_{12}^2\right]\nonumber \\ 
&&+ \frac{s_{14}}{ s_{34} s_{134}^2 } \left[ - 4 s_{12} s_{23} - 4 s_{12} s_{24}- 2 s_{12}^2 - 4 s_{23} s_{24}- 2 s_{23}^2- 2 s_{24}^2\right]\nonumber \\ 
&&- \frac{s_{14}}{ s_{234}} + \frac{s_{14}^2}{ s_{34}^2 s_{134}^2}  \left[- 4 s_{12} s_{23}- 4 s_{12} s_{24}- 2 s_{12}^2 - 4 s_{23} s_{24}- 2 s_{23}^2- 2 s_{24}^2\right]\nonumber \\ 
&&- \frac{1}{s_{12} s_{34} s_{234}}  \left[s_{13} s_{14}^2+ s_{13}^2 s_{14}- s_{13}^2 s_{23}+ s_{13}^3+ s_{14}^2 s_{23}+ s_{14}^3\right]\nonumber \\ 
&&- \frac{1}{s_{12} s_{34}}  \Big[s_{13} s_{14} - 3 s_{13} s_{23}- s_{13} s_{24}+ 2 s_{13}^2+ s_{14} s_{23}- s_{14} s_{24}\nonumber \\
&&+ 4 s_{14}^2+ s_{23} s_{24}+ s_{23}^2\Big]- \frac{1}{s_{12} s_{134}}  \Big[  s_{13} s_{24} - s_{14} s_{23}\nonumber \\ 
&&- 2 s_{14} s_{24}- s_{23} s_{24}+ s_{24}^2 \Big]- \frac{1}{s_{12}}  \left[s_{13}+ 2 s_{14} - s_{23}  - s_{24} \right]\nonumber \\ 
&&- \frac{s_{12}}{ s_{34} s_{134} s_{234}}  \left[  2 s_{14} s_{23} + 8 s_{14}^2 - 2 s_{12} s_{23} + 2 s_{12}^2 + 2 s_{23}^2\right]\nonumber \\ 
&&- \frac{s_{23}}{s_{34}^2  s_{234} } \Big[4 s_{13} s_{14} - 8 s_{12} s_{13} + 4 s_{13} s_{23} + 4 s_{13}^2 + 4 s_{14} s_{23} + 4 s_{12} s_{23}\nonumber \\ 
&& + 4 s_{12}^2\Big]+ \frac{s_{23}^2 }{s_{34}^2 s_{234}^2}  \left[ - 4 s_{13} s_{14} + 4 s_{12} s_{13} - 2 s_{13}^2 + 4s_{12} s_{14}  - 2 s_{14}^2 - 2 s_{12}^2\right]\nonumber \\ 
&&+ \frac{1}{s_{34}^2}  \Big[ 4 s_{13} s_{12}- 4 s_{13} s_{23}+ 4 s_{13} s_{24}- 2 s_{13}^2-4 s_{14} s_{12}  - 4 s_{14} s_{23}\nonumber \\ 
&& - 4 s_{14} s_{24} + 4 s_{12} s_{23}- 4 s_{12} s_{24}- 2 s_{12}^2- 2 s_{24}^2\Big]\nonumber \\ 
&&+ \frac{1}{s_{34} s_{134}}  \Big[  s_{12}s_{14} + 7 s_{14} s_{23}- s_{14} s_{24}+ 7 s_{14}^2+ s_{12} s_{23} + 7 s_{12} s_{24}\nonumber \\ 
&&+ 6 s_{12}^2+ 2 s_{23} s_{24} + s_{23}^2 + 3 s_{24}^2\Big]+ \frac{1}{s_{34} s_{234}}  \Big[-6  s_{12}s_{13}- 2 s_{13} s_{23} + 4 s_{13}^2\nonumber \\ 
&&  + 4 s_{14} s_{23}+ 6 s_{14}^2 + 5 s_{12}^2\Big]+ \frac{1}{s_{34}}  \left[3 s_{13}+ 2 s_{14}- 6 s_{12}- 6 s_{23} - s_{24}\right]\nonumber \\ 
&& + \frac{s_{23}}{ s_{24} s_{234}^2 } \left[ 2 s_{13} s_{14} - 2 s_{12}s_{13} + s_{13}^2- 2  s_{12}s_{14} + s_{14}^2 + s_{12}^2 \right]\nonumber \\
&&- \frac{1}{s_{24} s_{134} s_{234}}  \Big[  s_{12}s_{13} s_{14} - s_{13} s_{14} s_{23} - s_{13} s_{14}^2+  s_{12}s_{13} s_{23}+ 3  s_{12}^2s_{14}\nonumber \\ 
&&- 3  s_{12}s_{14}^2 + s_{14}^3- s_{12}^3 \Big]- \frac{1}{s_{24} s_{134}}  \left[   s_{12}s_{13}-s_{12} s_{14}  - s_{14}^2+ s_{12}^2 \right]\nonumber \\ 
&&- \frac{1}{s_{24} s_{234} } \Big[ 2 s_{12}s_{13} - 2 s_{13} s_{23} - s_{13}^2 - 3 s_{12}s_{14} - s_{14} s_{23}+ 4 s_{14}^2\nonumber \\ 
&&+ s_{12} s_{23} \Big]+ \frac{1}{s_{24} } \left[  2 s_{13}- 2 s_{14} + s_{34} \right]\nonumber \\ 
&&+ \frac{1}{s_{134}^2  }\left[- 2 s_{12} s_{23} - 2 s_{12} s_{24} - s_{12}^2 - 2 s_{23} s_{24} - s_{23}^2 - s_{24}^2\right]\nonumber \\ 
&&+ \frac{1}{s_{134} s_{234}}  \left[s_{13} s_{14} + 3s_{12} s_{13}  - s_{13}^2- 4 s_{12} s_{14} - s_{14}^2  - 3 s_{12}^2 \right]\nonumber \\ 
&&+ \frac{1}{s_{134} } \left[ - 3 s_{13}+ 6 s_{14} + 3 s_{23} - 2 s_{24}\right]+ 2\nonumber\\
&&+m_Qm_{\chi}\bigg[ -\frac{s_{12}}{s_{24} s_{134}}+\frac{3}{s_{34}}-\frac{s_{14}}{s_{34} s_{134}}+\frac{s_{14} s_{23}}{s_{24} s_{34}s_{134}}-\frac{2}{s_{134}}\nonumber\\
&& -\frac{s_{23}}{s_{24} s_{345}}+\frac{2 s_{23}}{s_{34} s_{345}}-\frac{1}{s_{345}}-\frac{4 s_{14}}{s_{12}s_{24}}+\frac{1}{s_{12}s_{34}}\left[s_{24}-s_{13}\right] \nonumber\\
&&+\frac{s_{14} s_{23}}{s_{12}s_{24} s_{34} }-\frac{s_{24}}{s_{12}s_{134}}-\frac{s_{14} }{s_{12}s_{34} s_{134} }\left[s_{23}+3s_{24} \right]\nonumber\\
&&-\frac{2s_{14}^2}{s_{12}s_{34}^2 s_{134} } \left[ s_{23} +s_{24} \right]+\frac{s_{13}}{s_{345}s_{12}}-\frac{s_{14} s_{23}}{s_{24} s_{345} s_{12}}\nonumber\\
&&-\frac{2 s_{13} s_{23}}{s_{34} s_{345} s_{12}}+\frac{2s_{23}^2}{s_{34}^2 s_{345} s_{12}} \left[s_{13} +s_{14}\right]\nonumber\\
&&+\frac{2}{ s_{12}s_{34}^2} \left[s_{13} s_{23}+s_{14} s_{23}-s_{13} s_{24}+s_{14} s_{24}\right]\bigg]\nonumber\\
&&+m_Q^2\bigg[\frac{s_{12}}{s_{24} s_{134}}-\frac{4}{s_{34}}+\frac{2}{s_{34} s_{134}} \left[2 s_{12}-s_{14}+s_{23}+3 s_{24}\right]+\frac{2 s_{23}}{s_{34}s_{345}}\nonumber\\
&&-\frac{1}{s_{134} s_{345}}\left[ s_{12}+s_{14}+s_{23}+s_{24}\right]+\frac{ s_{23}}{s_{24} s_{134} s_{345}}\left[s_{12}-s_{14}\right]\nonumber\\
&&-\frac{4 s_{23}}{s_{34} s_{134} s_{345}}\left[s_{23}+s_{14}\right]+\frac{1}{s_{345}}+\frac{1}{s_{34} s_{12}}\left[4 s_{14}-2 s_{23}-3 s_{24}\right]\nonumber\\
&&-\frac{2}{s_{34} s_{134}^2} \left[s_{12}^2+2  s_{12}s_{23}+2 s_{12}s_{24}+s_{23}^2+s_{24}^2+2 s_{23} s_{24}\right]\nonumber\\
&&+\frac{1}{s_{12} s_{134} }\left[2s_{23}-s_{24}\right]+\frac{1}{s_{12} s_{34} s_{134} }\left[2 s_{23}^2+5 s_{14} s_{23}+2 s_{24}^2-s_{14} s_{24}\right]\nonumber\\
&&+\frac{2}{s_{12} s_{34}^2 s_{134} } \left[s_{14}^2 s_{23}+ s_{14}^2 s_{24}\right]-\frac{2}{ s_{12}s_{34} s_{345}}\left[-2 s_{23}^2+s_{13} s_{23}-2 s_{14}s_{23}\right]\nonumber\\
&&-\frac{2 s_{23}^2}{ s_{12} s_{34}^2 s_{345}} \left[s_{13}+s_{14}\right]-\frac{2}{s_{12} s_{34}^2 }\left[ s_{13}s_{23}+s_{14} s_{23}-s_{13} s_{24}+s_{14} s_{24}\right]\nonumber\\
&&+\frac{2}{s_{12}}-\frac{2}{s_{12}^2}\left[s_{13}+s_{14}-s_{23}-s_{24}\right] -\frac{2 s_{23}}{s_{12} s_{345} }\nonumber\\
&&-\frac{2}{s_{12}^2 s_{34} } \left[s_{13}^2-2 s_{13}s_{23} +s_{14}^2+s_{23}^2+s_{24}^2-2 s_{14}s_{24}\right]\bigg]\nonumber\\
&&+m_Q^3m_{\chi}\bigg[ \frac{4}{s_{12}^2}-\frac{2}{s_{12} s_{34} s_{134}} \left[s_{23}+s_{24}\right]\bigg]+\frac{2m_Q^4}{s_{12} s_{34} s_{134}}\left[s_{23}+s_{24}\right]
\bigg\}\nonumber\\
&&+\order{\epsilon},\nonumber\\ \label{eq.E04ifm}
\eeqa
where $Q^2=-(p_1+p_3+p_4-p_2)^2$, $m_{\chi}=\sqrt{Q^2}$, $s_{134}=s_{13}+s_{14}+s_{34}$ and $s_{234}=s_{34}-s_{23}-s_{24}$.

The initial-final quark-gluon four-parton $\tilde{E}$ -type antenna is given by:
\beqa
\wt{E}_4^0(\Q{1},\q{3},\qb{4},\gli{2})&=&\frac{1}{\left(Q^2+m_Q^2\right)^2}\bigg\{ \frac{1}{s_{23} s_{24}}\Big[ -2 s_{12} s_{13}-2 s_{12} s_{14}-2 s_{12} s_{34}+2 s_{12}^2\nonumber\\
&&+2 s_{13} s_{34}+2 s_{13}^2+2 s_{14} s_{34}+2 s_{14}^2 \Big]+\frac{s_{23}}{s_{234}^2 s_{24}}\Big[-2 s_{12} s_{13}-2 s_{12} s_{14}\nonumber\\
&&+s_{12}^2+2 s_{13} s_{14}+s_{13}^2+s_{14}^2\Big]+\frac{s_{24}}{s_{23} s_{234}^2} \Big[-2 s_{12} s_{13}-2 s_{12} s_{14}+s_{12}^2\nonumber\\
&&+2 s_{13} s_{14}+s_{13}^2+s_{14}^2\Big]-\frac{1}{s_{234} s_{24}}\Big[4 s_{12} s_{13}+2 s_{12} s_{14}+s_{12} s_{23}-2 s_{12}^2\nonumber\\
&&-2 s_{13} s_{14}-s_{13} s_{23}-2s_{13}^2-s_{14} s_{23}\Big]-\frac{1}{s_{23} s_{234}}\Big[2 s_{12} s_{13}+4 s_{12} s_{14}\nonumber\\
&&+s_{12} s_{24}-2s_{12}^2-2 s_{13} s_{14}-s_{13} s_{24}-s_{14} s_{24}-2 s_{14}^2\Big]\nonumber\\   
&&-m_Q m_{\chi}\left[ \frac{2 s_{24}}{s_{23} s_{234}}+\frac{4 s_{234}}{s_{23} s_{24}}+\frac{2 s_{23}}{s_{234} s_{24}}+\frac{4}{s_{23}}+\frac{4}{s_{24}}\right] \bigg\}+\order{\epsilon},\nonumber\\ \label{eq.E04tifm}
\eeqa
with $Q^2$, $m_{\chi}$, $s_{134}$ and $s_{234}$ given as above. Both E-type antennae are normalised to the tree-level two-parton matrix element
\beq
\left| \cm_2(\tilde{\chi} g  \rightarrow \tilde{g})\right|^2=4 (1-\epsilon)\left[Q^2+m_Q^2\right]^2.
\eeq

\bibliography{bibliography}

\end{document}